\documentclass[12pt]{amsart}
\usepackage{graphics}
\usepackage{amsthm}
\usepackage{amssymb}
\usepackage{amsmath}
\usepackage{amsfonts}
\usepackage{epic}
\usepackage{curves}
\usepackage{bm}
\usepackage{amscd}

\newcommand{\baseRing}[1]{\ensuremath{\mathbb{#1}}}

\newcommand{\R}{\baseRing{R}}

\theoremstyle{plain}

\theoremstyle{definition}

\def\be{\begin{equation}}
\def\ee{\end{equation}} 
\title{The Particle Problem in Classical Gravity: A Historical Note on 1941 }

\author{Mariano Galvagno}

\author{Gast\'on Giribet}

\address{Departamento de F\'{\i}sica, Universidad de Buenos Aires.}

\begin{document}

%%%%%%%%%%%%%%%%%%%%%%%%%%%%%%%%%%%%%%%%%%%%%%%%%%%%%%%%%%%%%%%%%%%%%%%%

\begin{abstract}

This note is based on a relatively unknown paper of Albert Einstein published in 1941 in the {\it Revista de la Universidad Nacional de Tucum\'an}. That work can be regarded as the prequel of the one written in 1943 in collaboration with Wolfgang Pauli, where, along the same lines, a proof of the non-existence of certain non-singular solutions in Kaluza-Klein theory was given. More generally, disproving the existence of regular solutions in classical unified field theories became, specially after 1930, an important criterion leading Einstein's investigations on unified field theory. This is the context in which Einstein's paper of 1941 and its generalizations of 1943 and 1948 become important.

\end{abstract}

%%%%%%%%%%%%%%%%%%%%%%%%%%%%%%%%%%%%%%%%%%%%%%%%%%%%%%%%%%%%%%%%%%%%%%%%

\maketitle

\section{Introduction}

In 1941, Guido Fubini interceded in asking Albert Einstein to submit a contribution to the Argentinian journal {\it Revista de la Universidad Nacional de Tucum\'an}, which in the early 1940's was being founded by the Italian mathematician Alessandro Terracini (Terracini 1941; 1944). Einstein kindly agreed to submit an article to which he referred as ``{\it ein h\"ubscher Beweis station\"arer Gravitationsfelder}'' (Einstein, 1941c). In that article, a novel demonstration of the non-existence of regular static spherically symmetric solution in the general relativity was given.

Einstein's manuscript, which was described by Terracini in the Italian translation (Einstein, 1941b) as ``{\it una graziosa dimostrazione}'', proved in an ingenious way that no such regular solutions are admitted in the theory of gravitation $R_{\mu \nu}=0$ in asymptotically flat spacetime. The original title of the Einstein's handwritten draft in German was {\it Beweis der Nichtexistenz von Singularitaetsfreien Gravitationsfeldern mit nicht Verschwindender Gesamtmasse}, literal translation of the titles of (Einsein, 1941a; 1941d), and it is in possession of the {\it Biblioteca Mathematica} of the University of Turin. 

Einstein's proof acquires importance, not due to its particular application to the general theory of relativity, but due of its property of being suitable for generalizations to other theories of gravitation and unified fields. Generalizations of this sort are the one presented by Einstein and Pauli in 1943, in which the case of Kaluza-Klein theory was studied (Einstein \& Pauli, 1943), and the one presented by Papapetrou in 1948 for the case of the non-symmetric unified field theory (Papapetrou, 1948b); see also (Lichnerowicz, 1955).   

The paper of 1941 (Einstein, 1941a), relatively unknown, is cited for instance in (F\"olsing, 1993) and in other few articles dedicated to the study of the historical context; see for instance (van Dongen, 2002; Earman \& Eisenstaedt, 1999). It was Abraham Taub who reviewed Einstein's paper for the {\it Mathematical Review} in 1942 (Taub, 1942), and the translation of the article from the original in German was supervised by Terracini himself, who in a letter addressed to Einstein (Terracini, 1941) wrote: ``{\it Ihrem Wunsch nach, anstatt des deutschen Textes Ihres Aufsatzes, werden wir nebst der spanischen eine englische \"Ubersetzung ver\"offentlichen}, {expressing that both English and Spanish translations were going to be prepared}.'' The Spanish translation (Einstein, 1941d) and the English version (Einstein, 1941a) were published in the series A of the {\it Revista} (the section dedicated to mathematics and physics). In the subsequent editions, the journal continued publishing articles related to (unified) field theories; see for instance (Santal\'o, 1954; 1959).

The aim of the present note is not to explain the reasons that Einstein might have had to publish his {\it Beweis} in one of the first editions of an unknown Argentinian journal, but to explain the importance that such paper had for subsequent investigations of the author. We describe below the historical context, showing that the search for non-singular particle-like solutions to the gravitational field equations was of particular importance, and that paper was an important piece. In section 3, we summarize the result of Einstein's paper. In section 4, we comment on the generalizations.

\section{The historical context}

When studying the evolution of Einstein's ideas about the theory of fields in the second third of the 20th century, one inevitably arrives to the conclusion that the so-called particle problem in general relativity, meaning the problem of describing particles as pure field configurations, is inseparable of other two problems in the theory of the continuum, namely the problem of motion and that of the unification of the fields. Einstein's conviction about the fact that these three problems should be interrelated is particularly illustrated in the introduction of his famous paper written in collaboration with N. Rosen in 1935 (Einstein \& Rosen, 1935), where the authors state: ``[o]n the basis of the description of a particle without singularity one has the possibility of a logically more satisfactory treatment of the combined problem: the problem of the field and that of the motion coincide''. Indeed, the quest for regular solutions in classical unified field theory played a crucial role as a criterion leading Einstein's investigations. 

In the entire period 1919-1955 of Einstein's investigation, the particle problem was central and it mainly comprised two aspects of different degree of speculation: first, the representation of particles as regular solutions of a pure field theory, {\it i.e.} free of sources; second, the pious attempt to derive the atomistic nature\footnote{Even in what is regards to the quantum behavior of matter. This was based on the idea that the non-linearity of the gravitational (or unified) field theory would lead to mimic the conditions yielding to the quantum behavior of atoms.} of matter as a property encoded in the non-linearity of the theory of field. Here, we will refer to the former.

\subsubsection*{The period 1919-1935.} 

Focusing on the earliest periods, and trying to answer the question of what motivated Einstein's conviction that the particles should be represented as non-singular solutions of the gravitational (unified) field, we could start looking at the preliminary stages of the unification program, when Einstein, instead of thinking of particles as geons, ``had considered the possibility of interpreting the elementary electric and light quanta as certain singular points of generalized field equations''\footnote{Vizgin refers here to the epoch of the first attempts of finding a ``non-geometrized'' unified field theory.} (Vizgin, 1994). Vizgin, referring specifically to the research activities by 1923, pointed out that by then ``Einstein did not associate the problem of explaining the quantum behavior of [...] the electron with the maximum problem in the first stage of its solution, which consisted merely of proving that the field equations had non-singular static centrally symmetric solution that could be interpreted as the electron.'' In contrast, if one analyzes the ``principles'' that are enunciated in the subsequent work, cf. (Einstein \& Rosen, 1935), it becomes clear that such a picture changed later and the idea of particles as singularities of the field was abandoned. 

Regardless how much involved Einstein himself was in this particular aspect of the field theory by the beginning of the 1920's, the question about the existence of regular solutions of field equations representing particle-like objects was not ignored by the relativity community during that epoch. As explained by Vizgin, ``[i]t had become clear by the end of 1923 that none of the unified theories proposed since 1918 had such nonsingular solutions''. 

Later, after 1923, Einstein did dedicate attention to the problem of regular particle-like solutions, at least marginally. In a paper published in 1923 in collaboration with J. Grommer, he proved the non-existence of regular static spherically symmetric solutions in Kaluza's theory (Einstein \& Grommer, 1923). According to Vizgin, ``they tested the [Kaluza's theory] for the existence of centrally symmetric everywhere-regular solutions and concluded that they did not exist, this being, in Einstein's eyes, one indication of a merely formal nature of the unification''. 

\subsubsection*{The period 1935-1941.} 

Despite Einstein's categorical remarks on the failure of Kaluza's theory to describe regular particle-like solutions, his enthusiasm about the five-dimensional theory did not disappear immediately and, in fact, survived through the 1930's to conclude in 1943 with his paper with Pauli (Einstein \& Pauli, 1943). In an earlier period, Pauli and Einstein had already studied the validity of Weyl's theory based on the (non)-existence of static solutions in it\footnote{See, for instance, the letters addressed to Michele Besso, dated on December 12, 1919 (Einstein \& Besso, 1994); see also (Pauli, 1994).}. From this, we can appreciate to what extent the existence of solutions suitable to be interpreted as particle-like objects became a criterion for abandoning or not a tentative line of investigation. In fact, the particle problem played a central role after 1923, and continued like this through the 1920's and 1930's. The question arises as to whether or not Einstein's point of view about the description of particles as singular solutions of field equations changed after the 1930's. More specifically: what evidence we have, apart from the paper with Pauli, that the existence or non-existence of regular solutions was an important guideline in Einstein's search for a unified field theory in the period after 1935? This question finds a first answer in the following assertion: ``Every field theory, in our opinion, must therefore adhere to the fundamental principle that singularities of the field are to be excluded'', which can be found\footnote{Einstein commented this work (entitled ``the particle problem in the general theory of relativity'') in a letter addressed to Michel Besso, on February 16 of 1936, (Einstein \& Besso, 1994). In the letter to his friend, Einstein mentioned his hope of having achieved a first step towards a satisfactory theory of matter.} in (Einstein \& Rosen, 1935).  

Einstein's investigations on Kaluza-Klein theory during 1937-1943 also show in a clear way how important the search for singular solutions was in the study of unified field theory during the 1930s. We can quote an eloquent paragraph belonging to (Einstein \& Bergmann, 1938), in which the authors presented their ``generalization of Kaluza's theory''; they say: {``Many fruitless efforts to find a field representation of matter free from singularities based on this theory have convinced us, however, that such a solution does not exist''.} 

Even in the papers written in 1938 and 1939 in collaboration with L. Infeld and B. Hoffmann, dedicated to the study of the problem of motion in general relativity, where ``matter is represented as point singularities of the field'', Einstein suggested that ``this [singular aspect] may be due to our simplifying assumption that matter is represented by singularities, and it is possible that it would not be the case if we could represent matter in terms of a [unified] field theory from which singularities were excluded'' (Einstein, Infeld \& Hoffmann, 1938; Einstein \& Infeld, 1939). The analysis of the problem of motion proposed by Einstein, Infeld, and Hoffmann is often referred as an example of ``dualistic theory'' (Sen, 1968) because of its property of treating particles and fields as different entities. However, the conviction of the authors\footnote{See also the letter to Michele Besso, dated on August 16 of 1949, (Einstein \& Besso, 1994).} about the fact that this would correspond just to a ``provisional'' description is frequently omitted.

More emphasis appeared later, when Einstein admitted that the objection about the appearance of singular solutions in general relativity would be valid only if we consider this theory as a final theory of the whole field. Instead\footnote{According to Einstein's point of view reflected in his autobiographical notes (Einstein, 1949a).}, the field configuration corresponding to a particle should be considered not as a pure gravitational field, but as a solution of a unified field theory. According to this, a necessary requirement to eventually consider a field theory satisfactory would be its property of representing particles as solutions free of singularities in the whole space, but this can be only a requirement for the final theory and not for its imperfect approximations. This is the link between the problem of the description of the matter and the one of finding a unified field theory: Einstein preferred to admit explicitly the temporary inability to understand the structure of matter rather than to consider the image of particles as singularities of the gravitational field, which, according to his point of view, can only represent a temporary and approximate description (Einstein, 1953a).

\subsubsection*{The period 1941-1955.}

As we see, Einstein's interest in the particle problem continued after the 1930s. In 1941, he published his paper in the Argentinian journal (Einstein, 1941a), in which he proved the non-existence of static spherically symmetric solutions in general relativity. One year and a half later, he published his paper with Pauli, in which the proof was extended to the case of Kaluza-Klein five-dimensional cylindrical extension of four-dimensional spherically symmetric geometries. Previously, during the late 1930's, Einstein and collaborators Bergmann and Bargmann had been searching for such solutions in Kaluza-Klein theory with the intention to find a geometry suitable to be interpreted as the field configuration of a particle-like object. The quest for particle-like field solutions later continued, even beyond the moment when Einstein's enthusiasm about Kaluza's theory started to vanish. For instance, between 1948 and 1949, when Einstein's investigations on unified field theories were already entirely focused on the non-symmetric four-dimensional theories, he persisted in the search for the elusive non-singular solutions.

About the question of whether the existence of singularities should be admitted, Einstein maintained in 1953 ``that singularities must be excluded in a final theory of fields. It [did] not seem reasonable to introduce into a continuum theory points (or lines, etc\footnote{Clearly, referring to his previous investigations on the existence of regular solutions with axial symmetry in five-dimensional Kaluza-Klein theory.}) for which the field equations do not hold. Moreover, the introduction of singularities is equivalent to postulating boundary conditions (which are arbitrary from the point of view of the field equations\footnote{The emphasis on the arbitrariness is recurrent. He refers to a private communication with L. Silberstein where this aspect was explicitly pointed out and a particular example presented (Einstein \& Rosen, 1935).}) on surfaces which closely surround the singularities'' (Einstein, 1953a).

To emphasize the importance that the description of particles had within the framework of the classical field theory, we refer to (Tonnelat, 1955), where the author mentions, among the motivations for generalizing the general relativity, the fact that this theory ``separates quite radically the gravitational field from [...] the sources of field which conserves a phenomenological interpretation even when one talks about uncharged particles''. According to (Tonnelat, 1955), the separation between fields and particles was as important as the separation between different types of forces of nature.

\section{Einstein's paper in 1941}

Einstein's article, which we briefly describe below, begins with the proof of a theorem which can be called Einstein's ``theorem of infinitely close solutions''. This follows from the simple observation that if an arbitrary everywhere regular metric $g_{\mu \nu}$ and its infinitesimal deformation $g_{\mu \nu}+\delta g_{\mu \nu}$ are considered, then the associate deformation in the Ricci tensor (denoted by $\delta R_{\mu \nu}$) satisfies
\begin{equation}
\delta R_{\mu \nu} = - \nabla _{\rho } \delta \Gamma ^{\rho }_{\mu \nu} +\nabla _{\nu } \delta \Gamma ^{\rho }_{\mu \rho} + (\Gamma _{\sigma \nu}^{\rho} -\Gamma _{\nu \sigma}^{\rho}  ) \delta \Gamma ^{\sigma} _{\mu \rho}
\end{equation}
Then, by assuming the symmetry of the affine connection, $\Gamma _{\sigma \nu}^{\rho} =\Gamma _{\nu \sigma}^{\rho} $, and defining
\begin{equation}
U_{\mu \nu }^{\rho} = -\delta \Gamma^{\rho }_{\mu \nu }+ \frac 12 (\delta \Gamma ^{\sigma }_{\mu \sigma }\delta ^{\rho }_{\nu} +\delta \Gamma ^{\sigma }_{\nu \sigma }\delta ^{\rho }_{\mu} )
\end{equation}
one finds, with Einstein, that the following identity holds
\begin{equation}
\sqrt{-g} g^{\mu \nu } \delta R_{\mu \nu } = \partial _{\rho}  (\sqrt{-g} g^{\mu \nu } U^{\rho} _{\mu \nu})
\end{equation}
This implies that, for the cases where both the original and the varied metric satisfy the Einstein equation, i.e. $\delta R_{\mu \nu} = 0$, we get
\begin{equation}
\partial _{\rho } \left( \sqrt{-g} g^{\mu \nu } (-2\delta \Gamma^{\rho }_{\mu \nu }+ \delta \Gamma ^{\sigma }_{\mu \sigma }\delta ^{\rho }_{\nu} +\delta \Gamma ^{\sigma }_{\nu \sigma }\delta ^{\rho }_{\mu} ) \right) = 0  \label{cuatro}
\end{equation}
Then, we can expand around Minkowski spacetime as follows
\begin{equation}
g_{\mu \nu } = \eta _{\mu \nu } + h_{\mu \nu } \ \ , 
\end{equation}
and consider the Newtonian condition at the infinity, namely
\begin{equation}
h_{\mu \nu }= -\frac {2m}{r} \delta ^{\mu} _{\nu} \ \ \label{newt}.
\end{equation}
Then, we find that
\begin{equation}
\left( \sqrt{-g} g^{\mu \nu } (-2\delta \Gamma^{\rho }_{\mu \nu }+ \delta \Gamma ^{\sigma }_{\mu \sigma }\delta ^{\rho }_{\nu} +\delta \Gamma ^{\sigma }_{\nu \sigma }\delta ^{\rho }_{\mu} ) \right) = -4\delta m \partial ^{\rho } \left(r^{-1} \right)
\end{equation}
These equations ``hold for the infinitely close asymptotic equations. But equations (\ref{cuatro}) hold rigorously in the whole domain'', where we do not consider singularities. By integrating out the above equations over a domain defined by an hyper-surface $V \times \R $, where the factor $\R $ corresponds to the time direction and $V$ is a three-volume bounded by a sphere $S^2=\partial V$, we find that the Gauss' theorem implies
\begin{equation}
\partial _{\rho }\int _{V \times \R} \sqrt {-g} g^{\mu \nu} U^{\rho }_{\mu \nu} d^4x = 2\delta m  \int _{V \times \R } \nabla ^2  r^{-1} d^4x=0 
\end{equation}
since the surface integral must vanish according to (\ref{cuatro}). Here, we can consider an sphere $S ^2$ large enough in such a way that the Newtonian limit (\ref{newt}) results a good approximation. Hence, we get
\begin{equation}
\int _{V \times \R} \sqrt {-g} g^{\mu \nu} \delta R_{\mu \nu} d^4x= 8\pi \delta m \int _{\R } dt = 0
\end{equation}
and then
\begin{equation}
\delta m = 0  . \label{dfg}
\end{equation}

This implies that, under the assumption that the solutions of the field equations are regular everywhere, one is allowed to conclude that ``two infinitely close solutions without singularities have necessarily the same total mass $m$''. However, this confronts with the simple observation that if the metric $g_{\mu \nu} (x^{\rho})$ satisfies the field equations $R_{\mu \nu}=0$, then the same system is satisfied by the configuration $g_{\mu \nu} (\lambda x^{\rho})$ for any $\lambda \in \R $. Then, by identifying $\lambda ^{-1} = 1+\delta m/m$ in the spherically symmetric solution we get a contradiction; contradiction which, of course, ``disappears if we abandon the hypothesis on the non-existence of singularities.''

In this way, Einstein demonstrated the non-existence of gravitational fields with a non-vanishing total mass free from singularities in the general theory of relativity (Einstein, 1941a). 

\section{Generalizations to unified field theories}

Einstein's proof was generalized by Einstein and Pauli (Einstein \& Pauli, 1943) and by Papapetrou (Papapetrou, 1948b) to the cases of other (unified) field theories. About the former, let us make a remark: As it was signaled in (van Dongen, 2002), Einstein's assertion about the {\it scaling} map $x^{\rho } \to \lambda x^{\rho }$ as an application closed among a space of non-singular solutions is not general enough. Actually, this is the loophole in the argument that finally explains the existence of Kaluza-Klein non-singular particles despite the Einstein-Pauli generalization of Einstein's proof, which again follows from this class of scaling arguments. 

A few years after the publication of (Einstein, 1941a), Papapetrou (Papapetrou, 1948b) presented an alternative proof of Einstein's theorem which, while simpler, admits to be generalized without major difficulty to the case of the non-symmetric version of the gravitational theory. Papapetrou's proof follows from the so-called Tolman's energy theorem (see for instance (Tolman, 1987)) which states that the energy concentrated in a three-volume $V$ of a given static space-time is given by the formula 
\begin{equation}
{\mathcal U} = \int _{V} d^3x \sqrt {-g}(T^{1}_{1}+T^{2}_{2}+T^{3}_{3}-T^{0}_{0})
\end{equation}
where $T^{\mu}_{\nu}$ is the stress-tensor. Then, this allows to reobtain Einstein's conclusion in a rather simple way by noticing that the vanishing of $T_{\mu \nu}$ in the interior of $V$ implies that the mass $m$ of the asymptotically flat spherically symmetric solution vanishes as well. This observation leads to extend the Einstein's no-go theorem to the case of the non-symmetric Einstein-Straus theory, in which the tensor $T^{\mu}_{\nu}$ turns out to be replaced by an analogue$\ ^*T^{\mu}_{\nu}$, naturally defined by
\begin{equation}
\ ^*T_{\mu \nu} = \frac {1}{8\pi } \left( \ ^*R_{\mu \nu } -\frac 12 g_{\mu \nu } g^{\rho \gamma} \ ^*R_{\rho \gamma}   \right)
\end{equation}
where$\ ^*R_{\mu \nu}$ is the non-symmetric generalization of the Ricci tensor (Schr\"odinger, 1950). Then, by imposing the condition of vanishing divergence for the anti-symmetric tensor $\sqrt {-g} (g^{\mu \nu } - g^{\nu \mu })$, he is able to show that, in the weak-field approximation, the symmetric part of the metric tensor $g_{\mu \nu}$ satisfies the same equations that in the case of general relativity. Then, it is claimed that, even in the non-symmetric case, Einstein's proof holds straightforwardly\footnote{Through his deduction, Papapetrou discussed the ambiguity concerning the definition of the ``correct conservation laws'', arguing that in the static case such arbitrariness tends to disappear.}.

\section{Final remarks}

Einstein's paper published in the {\it Revista de la Universidad Nacional de Tucum\'an} in 1941 establishes the basis of subsequent works dedicated to disprove the existence of regular particle-like solutions to the equations of the unified field theories  (Einstein \& Pauli, 1943; Papapetrou, 1948b). The (non)-existence of such solutions was a guideline in the search of a satisfactory theory of the classical fields in the first half of the 20th century.

\medskip

\subsection*{Acknowledgments}

This work was supported by Universidad de Buenos Aires. The authors thank Ari Pakman and Gabriela Simonelli. We specially thank The Albert Einstein Archives, The Hebrew University of Jerusalem, Israel. G. Giribet thanks the Institute for Advanced Study. M. Galvagno thanks B.S. Portaro.

%\newpage

\end{document}